# APPLICATION OF COMPROMISING EVOLUTION IN MULTI-OBJECTIVE IMAGE ERROR CONCEALMENT

*Arash Broumand*

arash.broumand@alumni.ut.ac.ir

**ABSTRACT**

Numerous multi-objective optimization problems encounter with a number of fitness functions to be simultaneously optimized of which their mutual preferences are not inherently known. Suffering from the lack of underlying generative models, the existing convex optimization approaches may fail to derive the Pareto optimal solution for those problems in complicated domains such as image enhancement. In order to obviate such shortcomings, the *Compromising Evolution Method* is proposed in this report to modify the Simple Genetic Algorithm by utilizing the notion of compromise. The simulation results show the power of the proposed method solving multi-objective optimizations in a case study of image error concealment.

***Index Terms***— Compromise, Evolutionary Algorithm, Multi-objective Optimization, Image Error Concealment

## 1. INTRODUCTION

In practice, many decision making problems comprise a number of objectives to be achieved simultaneously. These objectives can be formulated by means of separate fitness functions, however, the preference of each function may be ambiguous due to the lack of a general governing model which prevents the derivation of a single description by a weighted fitness function that meets the general goal of optimization. The popular approach to these problems includes describing each objective by a number of constraints; since improving an objective is not possible without deteriorating other objectives in many cases, each objective determines uncompromising limits not to be exceeded by the others. In this way, only one fitness function is optimized and the other ones usually content with some empirically defined constraints. Suffering from the insufficiency of verifiable data, this approach may fail to achieve the optimality in the cases which the effect of the constrained objectives is not definitely understood. It is quite likely to occur especially in the problems that arise in dynamic environments where the state parameters may vary due to unpredictable causes. On the other hand, some hidden parameters may exist that affect the optimality of constrained solution. Thus, a constraint optimization approach may not be suitable confronting problems that involve latent parameters.

The characteristics of the optimal solution to a multi-objective problem is known as the notion of Pareto optimality [1]; the solution includes a set of feasible solutions which are superior to the others, even though none of them are optimal in regard to any single objective. In other words, the optimal solutions are those which are derived in a compromising manner, so that each fitness function gives up its best choice so that the other objectives achieve rather better solutions.

The *Compromising Evolution* is proposed in this report as a generic paradigm to consider the essence of Pareto optimality in the implementation of approaching it. To this end, the multi-objective problem is treated as a multi-sexual genetic algorithm (GA); in any generation, each of the solutions associated with an objective are considered to have a specific gender. A child is procreated by the combination of one parent from each gender, using the well-defined *crossover* and *mutation* operations in GA. A parent group makes a family of children of all genders. The children of a family are categorized in *commitment groups*, in which each gender has a member in it, for example, one brother and one sister in 2-gender realization. The concept of compromise is laid in the selection of survivals: the fitness of each child is evaluated by the fitness function corresponding to its gender. If a child dies, all members of its corresponding commitment group also die, no matter how good they are in terms of their fitness function. As the generations proceed, the survivals tend to the parents which compromise in a way that all of their children survive. Thus, the final solution is expected to relatively satisfy all objectives simultaneously.

As a case study, the proposed compromising evolutionary algorithm is employed to improve an image Error Concealment (EC) method known as *Sequential Best Range Matching* (SBRM) [2]. The aim of EC is to reconstruct the lost regions of a packet loss prone image, which appear as combinations of missed blocks of the size 8×8 due to the data structure of popular compression and transmission standards such as JPEG and MPEG. To each corrupted block of the image, the SBRM searches for the best range matching block so that their ranges comply with the most similarity in a particular sense such as MSE. Since the absence of intact regions affects the reliability of comparing criterion, the other important issue is to evaluate the reliability and take it into account in the searching procedure. To this end, the SBRM defines the *merit* of the image pixels in regard to the availability and the fidelity of reconstruction as the feature of reliability.

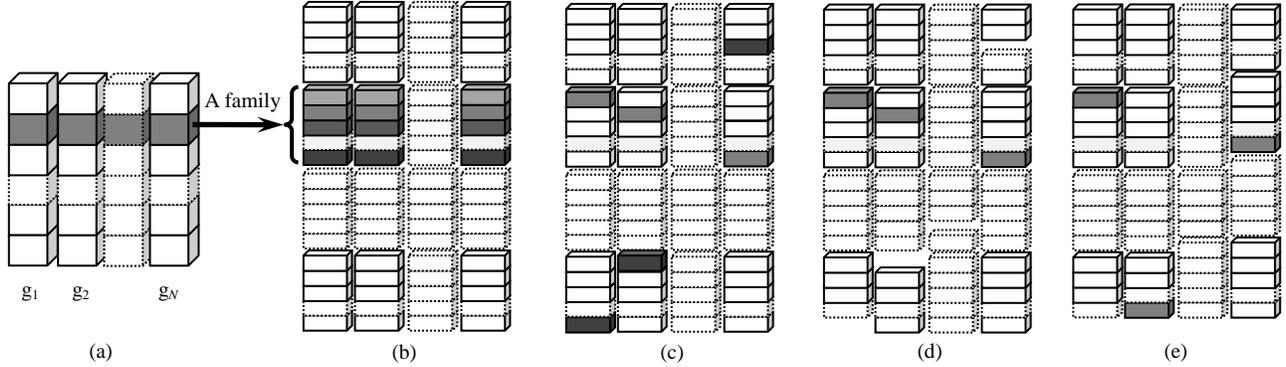

Fig.1. A scheme of generations, families, and compromising natural selection procedure, (a) the parent generation of genders $g_1$ to $g_N$, (b) the children, including commitment groups, (c) sorted children (d) The worst individual of $g_1$ causes the death of all its commitment group, (e) The same for $g_2$, and so on.

Although the mentioned two issues (the MSE and the reliability) are dissimilar in nature, the SBRM proposes a combination of them as a single fitness function, called *appraised MSE*, derived by massive experimental investigations. Yet the optimality of the proposed combined fitness function is not demonstrated. In order to obviate that deficiency, the problem should be treated as a multi-objective optimization. In this report, the Compromising Evolutionary algorithm is employed to simultaneously optimize the similarity and the reliability. The experimental results show that the proposed algorithm not only achieves the results provided by SBRM (i.e. it simply skips the experimental investigations necessary to determine the appropriate weights to combine fitness function), but also it considerably speeds up the process.

The rest of this report is organized as follows. The compromising evolutionary algorithm is proposed in the section. A brief overview on SBRM and the way the proposed method is applied to it is discussed in section 3. The experimental results are presented in section 4, and the conclusion remarks are made in section 5.

## 2. PROPOSED ALGORITHM

An *N*-objective optimization problem including $N$ fitness functions $\mathbf{F}(\mathbf{x}) = [F_1(\mathbf{x}), F_2(\mathbf{x}), ..., F_N(\mathbf{x})]$ of *D*-dimensional variables $\mathbf{x}$ is to be maximized. According to the concept of Pareto optimality, a Pareto optimal solution $\mathbf{x}^* \in R^D$ is a vector, so that there is no $\mathbf{x} \in R^D$ for which $\mathbf{F}(\mathbf{x})$ dominates $\mathbf{F}(\mathbf{x}^*)$, that is there is no $\mathbf{x}$ such that $\mathbf{F}_n(\mathbf{x}) \geq \mathbf{F}_n(\mathbf{x}^*)$ for all $n \in \{1, ..., N\}$. As described before, the proposed algorithm aims to acquire the concept of *compromise* in order to achieve the Pareto optimality.

Establishing the multiple-gender GA structure, each individual $\mathbf{x}$ of the population $X$ of $P$ members ($\mathbf{x} \in X^P$) is provided by an additional feature, the gender, which is stored in a vector $\mathbf{g}$. Initially, the gender of each individual in randomly determined so that the number of the population of all genders are identically *P/N*. Thus, $\mathbf{g}$ can be described as $\mathbf{g}^T = [\mathbf{g}_1^T, \mathbf{g}_2^T, ..., \mathbf{g}_N^T]$. To procreate the next generation, *P/N families* are established so that each individual of a gender population becomes a member of one family. Each family makes children by means of crossover and mutation operations so that the children are of all genders with identical number, that is each family includes *M* children of each gender which establish *M commitment groups*; each commitment group consists of one child from each gender. The new generation consists of *K* commitment groups as $C_k = \{\mathbf{x}_{k1}, \mathbf{x}_{k2}, ..., \mathbf{x}_{kN}\}$, $k = 1, ..., K = M \times P$.

Eventually, each gender population is evaluated and sorted according to its corresponding fitness function (Fig.1). The natural selection procedure is imposed in a sequential manner: the worst member of the first gender is determined by means of the first fitness function $F_1(\mathbf{x})$, and its corresponding commitment group is removed from the population. The same process is applied to the second gender and so on. The natural selection proceeds until the size of the overall survival population reaches *P*.

In this way, the number of those individuals increase which can procreate such children that all can survive; that is, the solutions which can satisfy all fitness functions survive as the generations proceed. The final solution is driven by a majority voting or the classification of the largest congregation of the last generation.

## 3. CASE STUDY: SBRM

In order to reconstruct the lost regions of an error prone image, the *Sequential Best Range Matching* method is an outperforming method in high packet loss ratios [2] in comparison with the existing error concealment techniques [3]-[7]. The reconstruction performs in an iterative manner; SBRM scans the image to find the corrupted blocks of the size *B×B* (Fig.3) in a raster scanning order. Each corrupted block gets one chance to be recovered as much as possible during each iteration. Let $\mathbf{x}$ be the vectorized pixel values of a basic block, so that the values of the lost pixels are set to zero. Similarly, $\mathbf{r}^x$ stands for the range block of $\mathbf{x}$ that contains the pixel values of an $R \times R$ block encompassing $\mathbf{x}$, where $R > B$. For each corrupted block, the algorithm searches for the best range matching block within an $A \times A$ *searching area* $\Lambda$ according to the minimum *appraised MSE* criterion, that is

$$e'_k = w_k e_k + (1 - w_k) \max_{\text{over } \Lambda} \{e_k\}, \qquad (4)$$

where $\Lambda$ is the set of all feasible blocks of the size $B \times B$ within the searching area encompassing the corrupted block, except for the corrupt itself, $e_k$ is the difference of the commonly available pixels of $\mathbf{r}^x$ and $\mathbf{r}^y$ in MSE sense, where $\mathbf{r}^y$ is the range of the $k^{\text{th}}$ block $\mathbf{y}_k \in \Lambda$, and the $w_k$ is the weighting factor as

$$w_k = \mathbf{m}^{x\,\mathrm{T}} \bullet \mathbf{m}_k^y \big/ nz(\mathbf{r}^x, \mathbf{r}_k^y) \qquad (5)$$

where T and • denote transport and inner product operations, $nz$ returns the number of commonly nonzero elements of the arguments, and $\mathbf{m}$ stands for the *merit vector* that indicates the reliability of each pixel of the range block; the elements corresponding to the intact pixels are initially set to 1, and the lost pixels are initially set to zero. As the iterations proceed, the merits of newly recovered pixels are set according to a decreasing function of the iteration number. The best matching block is selected as

$$\mathbf{y}^* = \mathbf{y}_k, \text{ so that } e'_k = \min_{\text{over } \Lambda} \{e'_l\}, \qquad (6)$$

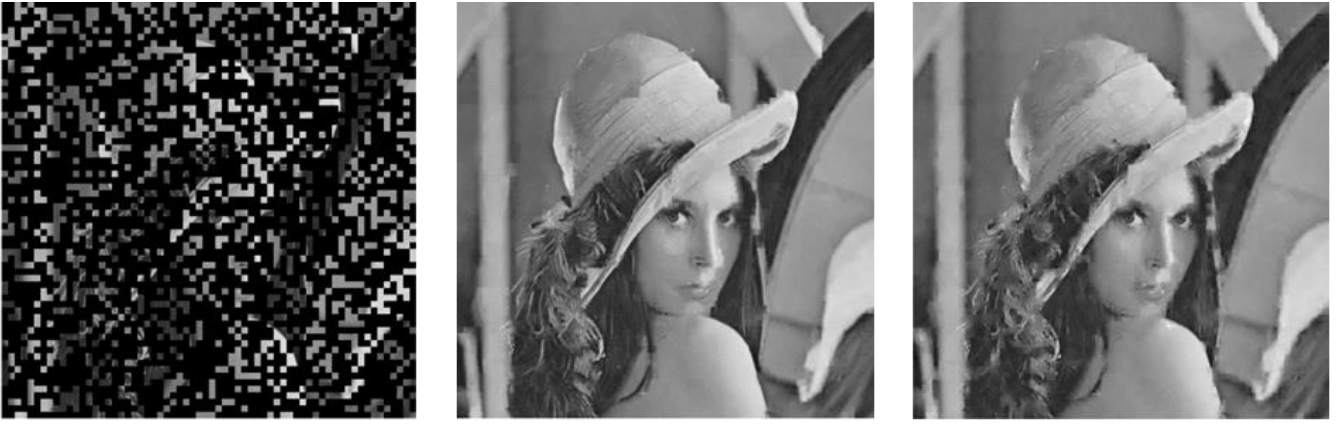

Fig.2. Left: Corrupted Lena, PSNR= 7.27 dB, with 70% packet loss. Center: Recovered by the original SBRM, PSNR= 23.91 dB, Right: Recovered by applying the Compromising Evolutionary on the SBRM, PSNR=23.53 dB, with 61% of time saving.

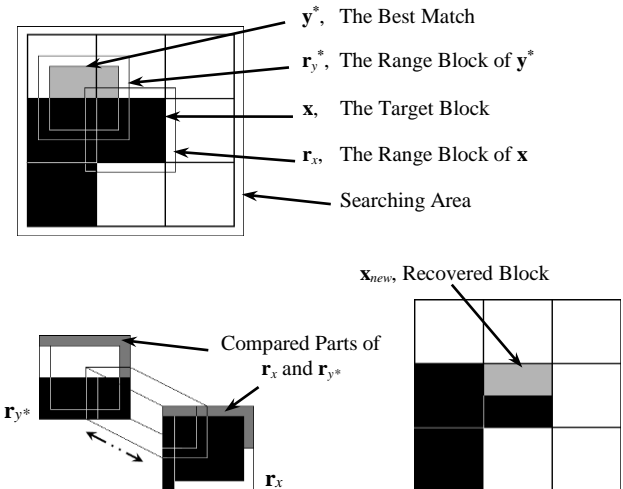

Fig.3. Top: Structure of the target and the best match blocks [2]. Bottom Left: Commonly available parts of ranges which are compared to evaluate the difference. Bottom Right: The recovered block.

The corrupted pixels are updated as follows

$$\mathbf{x}_{new} = \left[ x_{new,i} \right],\ x_{new,i} = \begin{cases} y_i^*, & \text{if } x_i \text{ is lost \& } y_i^* \text{ is available} \\ x_i, & \text{otherwise} \end{cases}. \quad (7)$$

The weighting factor is derived by huge experiments on some tested images; nevertheless it does not necessarily guarantee the performance of error concealment for any implementation. In order to avoid the case dependency of the weighting factor, the MSE criterion and the merit are considered as two separate fitness function; $e$, the difference of the available parts of two blocks in MSE sense, and the *mutual merit* $\mathbf{m}^{x\,\mathrm{T}} \bullet \mathbf{m}_k^y$ of them. The aim is to minimize $e$ and maximize the mutual merit simultaneously, in each trial of the reconstruction of a corrupted block. To this end, the proposed compromising evolutionary method is utilized as a bi-sexual algorithm.

## 4. EXPERIMENTAL RESULTS

The proposed compromising evolutionary method is applied to SBRM and the results are gathered for the standard Lena image in comparison with the existing error concealment methods in a broad range of packet loss ratios. The size of range blocks and searching areas used in the error concealment procedure are identical to [1] so that $B = 8$, $R = 10$ and $A = 3 \times 8$. The performance of the reconstruction process is evaluated in the sense of popular PSNR criterion as

$$PSNR = 10 \log \left( \frac{\max(\text{pixel value})^2}{\text{MSE of reconstructed image}} \right).$$

A perceptual comparison is made in Fig.2, where the performance of original SBRM and the compromising evolution driven SBRM are shown in the reconstruction of the corrupted Lena with 70% packet loss. The results show that applying the proposed method reduces the processing time of the original SBRM to the 39% in average, achieving approximately the performance which is still better than the other existing methods (Fig.4). The effect of the generation numbers of the evolutionary method is also shown in Fig.5.

## 5. CONCLUSION

A multi-sexual genetic algorithm based on the compromise in natural selection is proposed in this report as a generic approach to the multi-objective optimization problems, and its performance is shown by a promising application in the problem of error concealment in the quite complicated context of image [8].

The main advantage of the proposed approach is its inherent compatibility of implementation to the Pareto optimality concept. This advantage gives the opportunity of considering a huge number of objectives with no necessitation of deriving correlations among them. The proposed method can generally be used to avoid the affects of weak assumptions in the decision making procedures.

## ACKNOWLEDGEMENT

This report was a technical report prepared for the Bio-inspired Computations course instructed by late Prof. Caro Lucas. The author would like to thank Prof. Caro Lucas, and his former supervisor Dr. Alireza Nasiri Avanaki at the school of ECE, University of Tehran.

## REFERENCES

[1] Joanna Lis, and A. E. Eiben, "A Multi-sexual Genetic Algorithm for Multiobjective Optimization," *IEEE Int'l Conference on Evolutionary Computation*, 1997.

[2] Arash Baroumand, Alireza Nasiri Avanaki, "Sequential Best Neighborhood Matching: An Error Concealment Technique with Application in High Packet Loss Imgae Transmission," To be appeared in *IEEE ICASSP*, 2009.

[3] Gürkan Gür, Fatih Alagöz, and Mohammed AbdelHafez, "A Novel Error Concealment Method for Images Using Watermarking in Error-Prone Channels," *IEEE 16th Int'l Symposium on Personal, Indoor and Mobile Radio Communication*, 2005.

[4] Gürkan Gür, Yücel Altuğ, Emin Anarım, and Fatih Alagöz, "Image Error Concealment Using Watermarking with Subbands for Wireless Channels," *IEEE Comm. Letter*, vol. 11, no. 2, 2007.

[5] Onur G. Guleryuz, "Nonlinear Approximation Based Image Recovery Using Adaptive Sparse Reconstructions and Iterated Denoising," Part I and Part II, *IEEE Trans. on Image Processing*, vol. 15, no. 3, Mar. 2006.

[6] Zhou Wang, Yinglin Yu, and David Zhang, "Best Neighborhood Matching: An Information Loss Restoration Technique for Block-Based Image Coding Systems," *IEEE Trans. on Image Processing*, vol. 7, no. 7, July 1998.

[7] Xin Li, and Michael T. Orchard, "Novel Sequential Error-Concealment Techniques Using Orientation Adaptive Interpolation," *IEEE Trans. on Circuits And Systems for Video Technology*, vol. 12, no. 10, Oct. 2002.

[8] Fabrice Labeau, Claude Desset, Luc Vandendrope and Benoît Macq, "Performance of Linear Tools and Models for Error Detection and Concealment in Subband Image Transmission," *IEEE Trans. on Image Processing*, vol. 11, no. 5, May 2002.

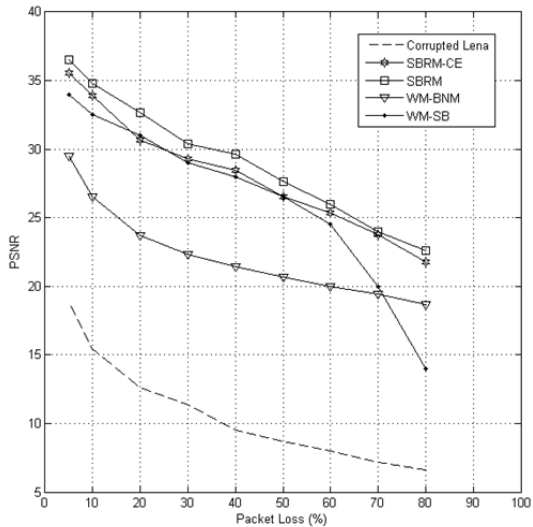

Fig.4. Performance of Compromising Evolutionary SBRM on Lena, a comparison to: Original SBRM [2], Watermarking-based BNM [3], and Watermarking with subbands [4].

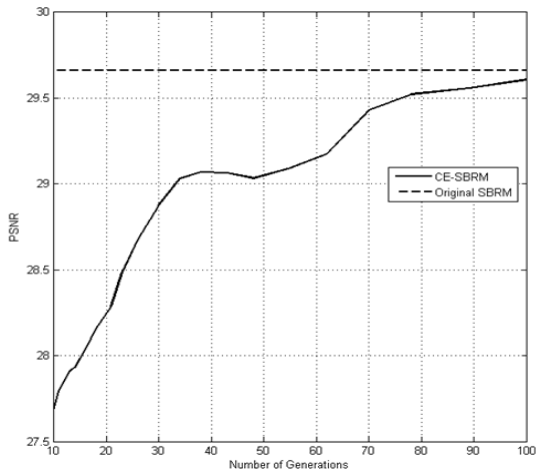

Fig.5. PSNR vs. the number of generations of the Compromising Evolutionary Algorithm, recovery of Corrupted Lena with 40% packet loss.